\documentclass[prl,10pt,twocolumn]{revtex4}

\usepackage[dvips]{epsfig}
\usepackage{float}
\usepackage{amsfonts}
\usepackage{amsmath}
\usepackage{amssymb}
\usepackage{enumerate}

\begin{document}

\title{Efficient input and output fiber coupling to a photonic crystal waveguide}

\author{Paul E. Barclay}
\email{pbarclay@caltech.edu}
\author{Kartik Srinivasan}
\author{Matthew Borselli}
\author{Oskar Painter}
\affiliation{Department of Applied Physics, California Institute of Technology, Pasadena, California 91125}
\date{\today}
\begin{abstract}
The efficiency of evanescent coupling between a silica optical fiber taper and a silicon photonic crystal waveguide is studied.  A high reflectivity mirror on the end of the photonic crystal waveguide is used to recollect, in the backwards propagating fiber mode, the optical power that is initially coupled into the photonic crystal waveguide.  An outcoupled power in the backward propagating fiber mode of $88\%$ of the input power is measured, corresponding to a lower bound on the coupler efficiency of $94\%$.
\end{abstract}

\maketitle

\setcounter{page}{1}
Evanescent coupling between optical fiber tapers\cite{ref:Knight} and photonic crystal waveguides (PCWG)\cite{ref:Barclay3,ref:OBrien2}  allows highly efficient and mode selective interfacing between fiber optics and planar photonic crystal devices.  Circumventing the intrinsic spatial and refractive index mismatch of fiber to PCWG end-fire coupling\cite{ref:Notomi}, this coupling scheme takes advantage of the strong dispersion and undercut geometry inherent to PC membrane structures to enable efficient power transfer between modes of a PCWG and a fiber taper.  The coherent interaction over the length of the coupling region between phase matched modes of each waveguide manifests in nearly complete resonant power transfer between waveguides, and by varying the position of the fiber taper it is possible to tune the coupling strength and phase matching frequency, allowing the dispersive and spatial properties of PCWG modes to be efficiently probed\cite{ref:Barclay4}.  

In previous work\cite{ref:Barclay3,ref:Barclay4}, the coupling efficiency of this technique was studied by analyzing the signal transmitted through the fiber taper as a function of wavelength and fiber taper position.  Although transmission measurements are sufficient to infer the efficiency of this coupling scheme, a measurement of power coupled into and then back out of a PCWG is of interest since it allows the lower bound of the input-output coupling efficiency to be directly established.  By studying light which is coupled into a PCWG, reflected by an end-mirror at the PCWG termination, and then coupled into the backward propagating fiber mode, we show in this letter that this coupling scheme has near unity efficiency.  

\begin{figure}[htb]
\begin{center}
\epsfig{figure=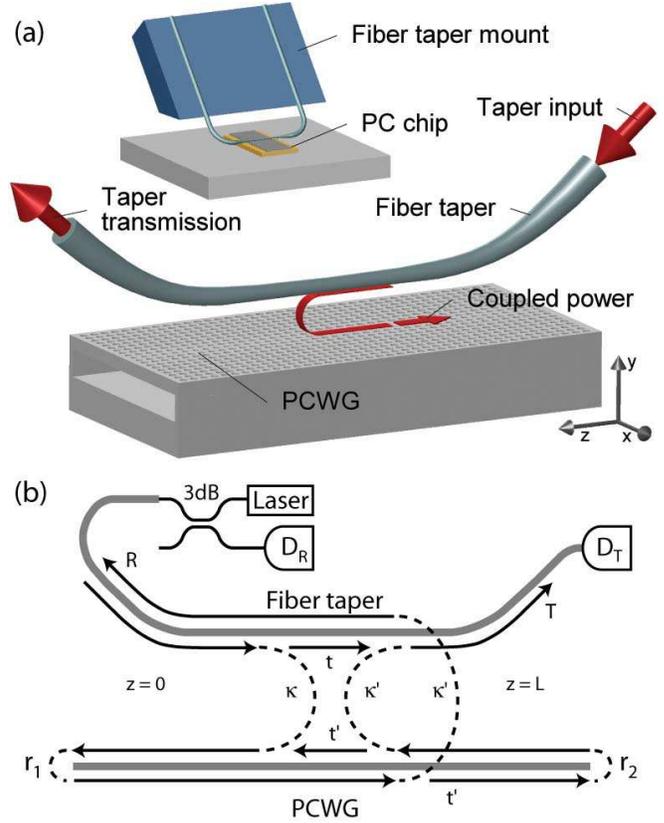, width=1.0\linewidth}
\caption{(a) Schematic of the coupling scheme. (b) Illustration of the contradirectional coupling process and the feedack within the PCWG caused by the reflectivities $r_{1,2}$ of the waveguide terminations.}
\label{fig:Couple_Scheme}
\end{center}
\end{figure} 

Figure \ref{fig:Couple_Scheme}(a) shows an illustration of the fiber taper evanescent coupling scheme.  A fiber taper, formed by simultaneously heating and stretching a standard single mode fiber until it has a minimum diameter of roughly a micron, is positioned above and parallel to an undercut planar PCWG.  The small diameter of the fiber taper allows the evanescent tail of its fundamental mode to interact with the PCWG, while the undercut geometry of the PCWG prevents the fiber taper from radiating into the PC substrate.   A schematic detailing the optical path within the fiber taper and the PCWG is shown in Figure \ref{fig:Couple_Scheme}(b).  To spectrally characterize the taper-PCWG coupler, an external cavity swept wavelength source with wavelength range $1565$ $\text{nm}$ - $1625$ $\text{nm}$ was connected to the fiber taper input.  At the fiber taper output, a photodetector ($\text{D}_{\text{T}}$) was connected to measure the transmitted power past the taper-PCWG coupling region.  An additional photodetector ($\text{D}_{\text{R}}$) was connected to the second 3dB coupler input to measure the light reflected by the PCWG. 

\begin{figure}[hb]
\begin{center}
\epsfig{figure=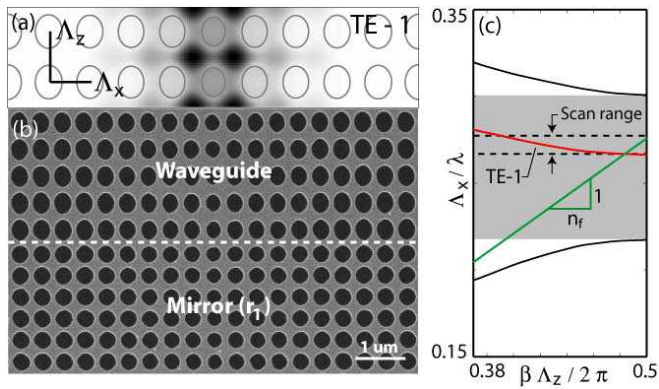, width=1.0\linewidth}
\caption{(a) Waveguide geometry and finite-difference time-domain (FDTD) calculated magnetic field profile ($B_y$) of the TE-1 mode. (b) SEM image of the high ($r_1$) reflectivity waveguide termination.  The PCWG has a transverse lattice constant $\Lambda_x = 415$ nm, a longitudinal lattice constant $\Lambda_z = 536$ nm, and length $L = 200\Lambda_z$. (c) Dispersion of the PCWG mode, and the band edges of the mirror termination for momentum along the waveguide axis ($\hat{z}$).}
\label{fig:SEM}
\end{center}
\end{figure}

The PCWGs used in this work were fabricated from a silicon-on-isulator (SOI) wafer. The air holes were patterned in the $300$ nm thick silicon (Si) layer using standard electron-beam lithography and dry etching techniques, after which the suspended membrane structure was created by selectively removing the $2$ $\mu$m thick underlying silicon dioxide layer using a hydorfluoric acid wet etch.  The PCWGs were formed by introducing a grade in hole radius along the transverse direction of a compressed square lattice of air holes\cite{ref:Barclay2}, as shown by Figure \ref{fig:SEM}(a-b).  Coupling in this case occurs between the fundamental linearly polarized fiber mode and the TE-1 PCWG mode, whose field profile is shown in Fig.\ 3(a).  This PCWG mode has a negative group velocity, resulting in \emph{contradirectional} coupling as depicted in Figure \ref{fig:Couple_Scheme}, and was designed to be integrated with recently demonstrated\cite{ref:Srinivasan4} high-Q cavites formed using a similar grading in hole radius.  

In previous studies of evanescent coupling to such PCWGs\cite{ref:Barclay4}, the waveguide termination consisted of a $2$ $\mu\text{m}$ region of undercut Si, followed by an $8$ $\mu\text{m}$ length of non-undercut SOI and a final air interface.  The interference from multiple reflections at the three interfaces within the waveguide termination resulted in a highly wavelength dependent reflection coefficient, making quantitative analysis of signals reflected by the PCWG difficult.  Here, the waveguide terminations were engineered to have either high or low \emph{broadband} reflectivities.  On one end of the PCWG, shown in Figure \ref{fig:SEM}(b), a photonic crystal mirror with high modal reflectivity ($r_1$) for the TE-1 PCWG mode was used.  This high reflectivity end mirror was formed by removing the lattice compression of the waveguide while maintaining the transverse grade in hole radius.  The change in lattice compression results in the TE-1 PCWG mode lying within the partial bandgap\cite{ref:Barclay2} of the high reflectivity end mirror section (Figure \ref{fig:SEM}(c)), while the transverse grade in hole radius reduces diffraction loss within the mirror.  On the opposite end of the PCWG, a poor reflector with low reflectivity ($r_2$) was realized by removing (over several lattice constants) the transverse grade in hole radius while keeping the central waveguide hole radius fixed, resulting in a loss of transverse waveguiding within the end mirror section and allowing the TE-1 PCWG mode to diffract into the unguided bulk modes of the PC.

In the absence of reflections due to the PCWG termination (i.e.\ $r_{1,2} = 0$), the fiber-PCWG junction can be characterized by 
\begin{equation}\label{eq:coupling_matrix}
  \left[
  \begin{array}{c}
    a^+_{\text{F}}(L) \\
    a^-_{\text{PC}}(0) 
  \end{array}
  \right]
     = 
  \left[
  \begin{array}{cc}
    t & \kappa' \\
    \kappa & t'
  \end{array}
  \right]
  \left[
  \begin{array}{c}
    a^+_{\text{F}}(0) \\
    a^-_{\text{PC}}(L)
  \end{array}
  \right],
\end{equation}
\noindent where $\kappa$ and  $\kappa'$ are coupling coefficients, $t$ and $t'$ are transmission coefficients, and $a^+_{\text{F}}(z)$ and $a^-_{\text{PC}}(z)$ are the amplitudes of the forward propagating fundamental fiber taper mode and the backward propagating TE-1 PCWG mode, respectively.  The coupling region extends along the $z$-axis, with $z=0$ corresponding to the input and $z=L$ to the output of the coupler.  As illustrated by Figure \ref{fig:Couple_Scheme}(b), non-zero $r_{1,2}$ introduce feedback into the system.   This allows input light coupled from the forward propagating fundamental fiber taper mode into the TE-1 PCWG mode to be partially reflected by the PCWG termination, and subsequently coupled to the backward propagating fundamental fiber taper mode.   In the presence of this feedback within the PCWG, the normalized reflected and transmitted powers in the fiber taper are given by\cite{ref:Haus_Book}
\begin{align}
T = \left| a^+_{\text{F}}(L)\right|^2 &= \left|t + \frac{\kappa \kappa' t' r_1 r_2}{1 -  r_1 t' r_2 t'}\right|^2,\label{eq:trans} \\
R = \left|a^-_{\text{F}}(0)\right|^2 &= \left|\frac{\kappa \kappa' r_1}{1 -  r_1 t' r_2 t'}\right|^2,\label{eq:refl} 
\end{align}
\noindent for $a^+_{\text{F}}(0) = 1$ and $a^-_{\text{PC}}(L) = 0$.  By measuring $T$ and $R$, and considering Eqs. (\ref{eq:trans},\ref{eq:refl}) we can determine the efficiency of the fiber-PCWG coupling as measured by $|\kappa\kappa'|$.  

\begin{figure}[bh]
\begin{center}
\epsfig{figure=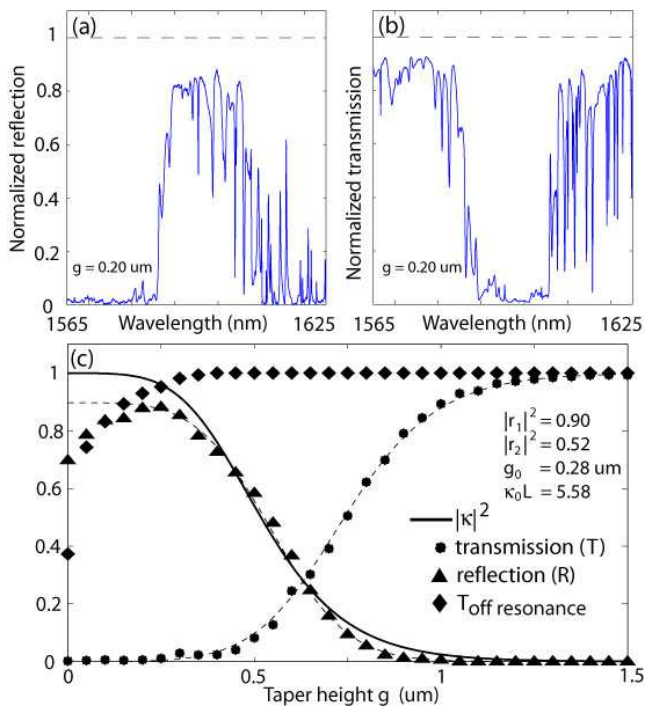, width=0.98\linewidth}
\caption{(a) Reflection and (b) transmission of the fiber taper as a function of wavelength for a taper height of $0.20$ $\mu$m.  Both signals were normalized to the taper transmission with the PCWG absent. (c) Measured taper transmission minimum, reflection maximum, and off-resonant transmisson as a function of taper height.  Also shown are fits to the data, and the resulting predicted coupler efficiency, $|\kappa|^2$.  Asymmetry in the fiber taper loss about the coupling region was taken into account by repeating the measurements with the direction of propagation through the taper and the orientation of the PCWG sample reversed; the geometric mean of the values obtained from the two orientations takes any asymmetry into account.}
\label{fig:Data}
\end{center}
\end{figure}   

In the measurements presented here a fiber taper with $2$ $\mu$m diameter was used to probe the PCWG of Figure \ref{fig:SEM}(b).  Figures \ref{fig:Data}(a) and \ref{fig:Data}(b) show $T$ and $R$ as a function of wavelength when the taper is aligned with the PCWG at a height of $g = 0.20$ $\mu$m, and indicate that the phase-matched wavelength is $\lambda \sim 1598 \pm 5$ nm.  As described in Ref.\ \cite{ref:Barclay4}, coupling to the guided TE-1 mode of the PCWG was confirmed by studying the coupling dependence upon polarization, lateral taper displacement, and fiber taper diameter.  Figure \ref{fig:Data}(c) shows $T$ and $R$ at the resonant wavelength of the PCWG nearest the phase-matching condition (the minima and maxima in $T(\lambda)$ and $R(\lambda)$), as a function of taper height $g$ above the PCWG.  Included in Figure \ref{fig:Data}(c) is the off-resonant (away from phase-matching) transmission, $T_{\text{off}}$, through the fiber.  A maximum normalized reflected power $R_{\text{max}} \sim 0.88$ was measured at a height of $g \sim 0.25$ $\mu\text{m}$, where the corresponding transmission was $T < 0.01$.  As can be seen in the wavelength scans of Figure \ref{fig:Data}(a), at this taper height the Fabry-Perot resonances due to multiple reflections from the end mirrors of the PCWG are suppressed for wavelengths within the coupler bandwidth due to strong coupling to the fiber taper.  Ignoring for the moment the effects of multiple reflections in eq.\ \ref{eq:refl}, the maximum optical power coupling efficiency is then $|\kappa\kappa'| = \sqrt{R_{\text{max}}}/|r_1|$, where the square root dependence upon $R_{\text{max}}$ is a result of the light passing through the coupler twice in returning to the backward propagating fiber mode.  Assuming the high reflectivity mirror is perfect (unity reflection), for the measured $R_{\text{max}}=0.88$ this implies $94\%$ coupling of optical power from taper to PC waveguide (and vice-versa).      

By further comparison of $T(g)$ and $R(g)$ with the model given above, this time including feedback within the PCWG, we find that the high reflectivity PCWG end-mirror is imperfect and that $R_{\text{max}}$ is in fact limited by mirror reflectivity, not the efficiency of the coupling junction. For this comparison we take the elements of the coupling matrix to satisfy the relations $|t|^2 + |\kappa|^2 = 1$, $\kappa' = \kappa$, and $t' = t^*$ of an ideal (lossless) coupling junction\cite{ref:Haus_Book}.  For the phase matched contradirectional coupling considered here, the dependence of $\kappa$ on $g$ can be approximated by $|\kappa(g)| = \tanh\left[\kappa_{\perp,0}\exp(-g/g_0)L\right]$\cite{ref:Yeh2}, where $\kappa_{\perp,0}$ and $g_0$ are constants.  Substituting these relations into eqs.\ (\ref{eq:trans}) and (\ref{eq:refl}), $T$ and $R$ can be fit to the experimental data with $r_{1,2}$, $\kappa_{\perp,0}$ and $g_0$ as free parameters\cite{ref:phase_footnote}.  Note that waveguide loss can also be included in the model, however it is found to be small compared to the mirror loss and will be studied in detail elsewhere.  Figure \ref{fig:Data}(c) shows the fits to the data for $g \ge 0.20$ $\mu\text{m}$; for $g < 0.20$ $\mu\text{m}$ non-resonant scattering loss is no longer negligible and the ideal coupling junction model breaks down.  Note that the off-resonant scattering loss is observed to effect the reflected signal at a smaller taper-PCWG gap than for the transmitted signal, indicating that within the coupler bandwidth $1-T_{\text{off}}$ is an overestimate of the amount of power scattered into radiation modes.  From the fits in Figure \ref{fig:Data}(c), the PCWG end-mirror reflectivies are estimated to be $|r_1|^2 = 0.90$ and $|r_2|^2 = 0.52$, and the optical power coupling efficiency ($|\kappa|^2$) occurring at $R_{\text{max}}$ ($g=0.25$ $\mu$m) is approximately $97\%$.

In conclusion, these results indicate that this coupling scheme allows highly efficient input and output coupling between photonic crystal devices and fiber tapers, and is promising for efficient probing of future integrated planar PC devices.

KS would like to thank the Hertz foundation and MB the Moore foundation for financial support.


\begin{thebibliography}{10}
\bibitem{ref:phase_footnote}
We take $r_1 t' r_2 t' /|r_1 t' r_2 t'| = 1$, corresponding to a resonant
  condition within the PCWG.
\end{thebibliography}
\end{document}